\documentclass[runningheads]{svmult}

\usepackage{makeidx}   
\usepackage{graphicx}  
\usepackage{subeqnar}  
\usepackage{multicol}  
\usepackage{physprbb}  
\makeindex             

\def\psim{\lower.5ex\hbox{$\; \buildrel \propto \over\sim \;$}}
\def\gtrsim{\lower.5ex\hbox{$\; \buildrel > \over\sim \;$}}
\def\lesssim{\lower.5ex\hbox{$\; \buildrel < \over\sim \;$}}

\begin{document}
\title*{Ultra-High Energy Cosmic Rays and Neutron-Decay Halos from Gamma Ray Bursts }
\toctitle{Ultra-High Energy Cosmic Rays and Neutron-Decay Halos from Gamma Ray Bursts}
%
%
\titlerunning{Ultra-High Energy Cosmic Rays from GRBs}
\author{C. D. Dermer}
\authorrunning{C. D. Dermer}

\institute{Code 7653, Naval Research Laboratory, 4555 Overlook Ave., SW, Washington, DC 20375-5352 USA}

\maketitle    

\begin{abstract}
Simple arguments concerning power and acceleration efficiency show that ultra-high energy cosmic rays (UHECRS) with energies $\gtrsim 10^{19}$ eV could originate from GRBs.  Neutrons formed through photo-pion production processes in GRB blast waves leave the acceleration site and travel through intergalactic space, where they decay and inject a very energetic proton and electron component into intergalactic space. The neutron-decay protons form a component of the UHECRs, whereas the neutron-decay electrons produce optical/X-ray synchrotron and gamma radiation from Compton-scattered background radiation. A significant fraction of galaxies with GRB activity should be surrounded by neutron-decay halos of characteristic size $\sim 100$ kpc. 

\end{abstract}

\section{Introduction}

Gamma-ray bursts produce enough power within the Greisen-Zatsepin-Kuzmin (GZK) photopion production radius to power the UHECRs [1,2,3]. Stochastic gyroresonant acceleration of protons and ions by turbulence generated in relativistic blast waves can accelerate particles to ultra-high energies [4]. Energetic neutrons are formed by photopion interactions of accelerated hadrons with nonthermal synchrotron radiation in GRB blast waves. The neutrons travel through intergalactic space and decay, and the neutron-decay electrons form synchrotron and Compton halos around galaxies with GRB activity. The discovery of neutron-decay halos around galaxies with vigorous star-forming activity will provide strong evidence for a GRB origin of the UHECRs [3].  

\section{GRB Origin of UHECRs: Power and Acceleration}

As a consequence of Beppo-SAX results, we now know that GRBs are extragalactic and originate from sources with a broad distribution of redshifts and mean redshift $\bar z \approx 1$. Beppo-SAX has a much smaller field-of-view than BATSE, but triggers on nearly the same sample of long-duration ($t_{50}\gtrsim 1$ s) GRBs. If UHECRs originate from GRBs, then the product of the UHECR energy density $u_{UH}$ and the characteristic source volume $V$ is equal to the product of the GRB power $L_{GRB}$, the loss time from the source volume, and the efficiency $\epsilon$ to convert GRB energy into UHECRs. For protons with energies $\gtrsim 10^{20}$ eV, the GZK radius is $\sim 140$ Mpc [5]. Thus the loss time $t_{p\gamma} \cong 140 $ Mpc/c $\cong 1.4\times 10^{16}$ s. Hence $u_{UH} \cong \epsilon f L_{GRB}t_{p\gamma}/V$, where $f$ is a factor that takes into account present day star-formation activity compared with that occurring at $\bar z$.  If $\bar z = 1$, then $f \cong 1/6$.

Let $\bar d = 10^{28}d_{28}$ cm represent the average luminosity distance to observed GRBs, so that $V \cong 4\pi \bar d^3/3$. The power of GRBs into the volume $V$ is given by the typical GRB energy $E_{GRB}$ multiplied by the GRB rate. BATSE is sensitive to GRBs with peak fluxes $\phi \gtrsim 10^{-7}\phi_{-7}$ ergs cm$^{-2}$ s$^{-1}$. The observed mean duration of the long duration GRBs is $t_{dur} = 30 t_{30}$ s. Thus $E_{GRB} \approx 4\pi \bar d^2 \cdot 10^{-7}\phi_{-7}\cdot 30 t_{30} (1+\bar z)/(0.1\eta_{-1}) \approx 4\times 10^{52}d_{28}^2 \phi_{-7} t_{30} (1+\bar z)/\eta_{-1}$ ergs, where $\eta = 0.1\eta_{-1}$ is the efficiency for transforming the GRB explosion energy into $\gamma$ rays in the Beppo-SAX and BATSE energy bands. The long-duration GRBs occur at a rate of $\approx 1/(t_{day}$ day), with $t_{day} \approx 1$, so that $L_{GRB} \approx 4\times 10^{47}d_{28}^2 \phi_{-7} t_{30} (1+\bar z)^2/(\eta_{-1}t_{day})$ ergs s$^{-1}$. We therefore find that 
\begin{equation}
u_{UH}\;({\rm ergs~ cm}^{-3}) \approx 1.5\times 10^{-21} {k\epsilon f \phi_{-7} t_{30} (1+\bar z)^2\over \eta_{-1} t_{day} d_{28}}\; .
\end{equation}
The factor $k$ represents the energy released by the dirty and clean fireballs which do not trigger the BATSE detector. Detailed calculations within the context of the external shock model show that $k \approx 3$ [5].

Observations show that $u_{UH} \cong  10^{-20}$ and $2\times 10^{-21}$ ergs cm$^{-3}$ for cosmic rays with $E \gtrsim 10^{19}$ eV and $10^{20}$ eV, respectively. (For protons with energies $\gtrsim 10^{19}$ eV, $t_{p\gamma}\cong 1000$ Mpc/c.) If an efficient mechanism for converting the energy of the relativistic outflows into UHECRs exists, then there is sufficient power in the sources of GRBs to power the UHECRs.  Detailed calculations [3,6,7] verify this result. 

Particle acceleration in GRB blast waves must satisfy the Hillas [8] condition for UHECR production, which requires that the Larmor radius be smaller than the characteristic size of the acceleration region. For GRB blast waves, this size is the blast-wave width.  Hence the particle Larmor radius $r_{\rm L} = (A m_pc^2/ZeB)$ $(\gamma_{max}/\Gamma ) < \Delta^\prime = f_\Delta x/\Gamma$, where $\gamma_{max}$ is the maximum particle Lorentz factor measured in the explosion frame, $\Gamma = 300\Gamma_{300}$ is the blast wave Lorentz factor, $\Delta^\prime$ is the comoving blast wave width, $f_\Delta \cong 1/12$ from hydrodynamics, and $x = 10^{16} x_{16}$ cm is the location of the blast wave from the explosion center. The blast-wave magnetic field $B \cong \sqrt{32 \pi e_B n_{ISM} m_pc^2}\Gamma \cong 0.4 \sqrt{e_B n_{ISM}}\Gamma$ G is defined by a magnetic-field parameter $e_B(< 1)$, and the term $n_{ISM}$ is the particle density of the surrounding medium. Thus
\begin{equation}
E_{max} = A m_pc^2 \gamma_{max} = ZeB f_\Delta x \simeq 3\times 10^{19}Z \sqrt{e_B n_{ISM}} ({f_\Delta\over 1/12})\; x_{16} \Gamma_{300} \;{\rm eV}.
\end{equation}
A wide range of parameter values can satisfy the Hillas condition for accelerating UHECRs by stochastic acceleration through gyroresonant interactions with MHD turbulence in the blast wave fluid [4,9]. The Alfv\'en speed $v_{\rm A}$ in the relativistic shocked fluid is also relativistic (naively using the nonrelativistic expression gives $v_{\rm A}/c \cong \sqrt{2e_B\Gamma}$, resulting in an acceleration rate that is much more rapid for second-order than for first-order processes [9,10].

\section{Neutron-Decay Halos}

Accelerated protons and ions interacting with nonthermal synchrotron radiation in the blast wave will produce neutrons through the process $p+\gamma \rightarrow n + \pi^+$. The neutrons, unbound by the magnetic field in the blast wave, leave the acceleration site with Lorentz factors $\gamma_n = 10^{10}\gamma_{10}$, with $0.1 \lesssim \gamma_{10} \lesssim 100$. The neutrons decay on a timescale $\gamma_n t_n \simeq 3\times 10^5 \gamma_{10}$ yr, where the neutron $\beta$-decay lifetime $t_n \cong 900$ s.  The neutrons travel a characteristic distance $\lambda_n \simeq 90\gamma_{10}$ kpc before they decay and inject highly relativistic electrons and protons into intergalactic space. Approximately 1\% of the energy of a GRB explosion with $10^{54}E_{54}$ ergs is deposited into highly relativistic neutrons when $E_{54} \approx 1$ [3]. The neutron-decay electron halo surrounding a galaxy from a single GRB reaches a maximum power of $L_{halo} \approx 0.01\times {\cal F} 10^{54}E_{54} (m_e/m_p)/(\gamma_n t_n) \approx 10^{36}E_{54}{\cal F}/\gamma_{10}$ ergs s$^{-1}$. Detailed calculations show that ${\cal F} \approx 0.1$ [3]. The neutron-decay protons become part of the UHECRs. GRB explosions with $E_{54} \gtrsim 0.2$ occur at a rate of about once every 5 Myrs per L$^*$ galaxy, implying that $\sim (5$-10)$\gamma_{10}$\% of L$^*$ galaxies should display a neutron-decay halo at maximum power. If GRB emission is beamed, a larger fraction of galaxies will display proportionally weaker halos.

The beta-decay electrons radiate nonthermal synchrotron emission and Compton scatter CMB photons to high energies. The maximum synchrotron frequency is $\nu \sim 3\times 10^{20}B(\mu{\rm G})\gamma_{10}^2$ Hz, where $B(\mu{\rm G})$ is the mean magnetic field in the region surrounding the galaxy in $\mu$G. The halo will display a cooling synchrotron spectrum at optical and soft X-ray energies. The electromagnetic cascade formed by the Compton-scattered $\gamma$ rays terminates when the $\gamma$ rays are no longer energetic enough to pair produce with the diffuse radiation fields. The relative intensity of the synchrotron and Compton components depends on the magnitude of $B(\mu{\rm G})$. The best prospect for discovering neutron-decay halos is by optical observations of field galaxies that display active star formation [3].

We also note that the emission of nonthermal synchrotron and Compton radiation from photopion processes by UHECRs traveling through intergalactic space will produce a nonthermal component of the diffuse radiation background, irrespective of the sources of UHECRs.

\end{document}